%% file: PTF13ast-revised-alex.tex
\documentclass[nature]{article}
\usepackage[margin=1in]{geometry}
\usepackage{graphicx}
\usepackage{epsfig}
\usepackage{fullpage}
\usepackage[super,sort&compress]{natbib}
\usepackage{epstopdf}
\usepackage{caption}
\newcommand{\spacing}[1]{\renewcommand{\baselinestretch}{#1}\large\normalsize}
\spacing{2}
\def\@maketitle{%
  \newpage\spacing{1}\setlength{\parskip}{12pt}%
    {\Large\bfseries\noindent\sloppy \textsf{\@title} \par}%
    {\noindent\sloppy \@author}%
}
\newenvironment{affiliations}{%
    \setcounter{enumi}{1}%
    \setlength{\parindent}{0in}%
    \slshape\sloppy%
    \begin{list}{\upshape$^{\arabic{enumi}}$}{%
        \usecounter{enumi}%
        \setlength{\leftmargin}{0in}%
        \setlength{\topsep}{0in}%
       \setlength{\labelsep}{0in}%
        \setlength{\labelwidth}{0in}%
        \setlength{\listparindent}{0in}%
        \setlength{\itemsep}{0ex}%
        \setlength{\parsep}{0in}%
        }
    }{\end{list}\par\vspace{12pt}}

\renewenvironment{abstract}{%
    \setlength{\parindent}{0in}%
    \setlength{\parskip}{0in}%
    \bfseries%
    }{\par\vspace{-6pt}}

\title{A Wolf-Rayet-like progenitor of supernova SN 2013cu from spectral observations of a wind}
\author{Avishay Gal-Yam$^{1,2}$, I. Arcavi$^{1}$ , E. O. Ofek$^{1}$, S. Ben-Ami$^{1}$, S. B. Cenko$^{3}$, \\
M. M. Kasliwal$^{4}$, Y. Cao$^{5}$, O.Yaron$^{1}$, D. Tal$^{1}$, J. M. Silverman$^{6}$, A. Horesh$^{5}$,\\ 
A. De Cia$^{1}$, F. Taddia$^{7}$, J. Sollerman$^{7}$, D. Perley$^{5}$, P. M. Vreeswijk$^{1}$, \\
S. R. Kulkarni$^{5}$, P. E. Nugent$^{8}$, A. V. Filippenko$^{9}$, \& J. C. Wheeler$^{6}$}

\begin{document}
\setcounter{page}{0}
\maketitle
\begin{affiliations}
 \item Department of Particle Physics and Astrophysics, Weizmann Institute of Science, Rehovot 76100, Israel.       
 \item Kimmel Investigator.
 \item Astrophysics Science Division, NASA Goddard Space Flight Center, Mail Code 661, Greenbelt, MD 20771, USA.
 \item Observatories of the Carnegie Institution for Science, 813 Santa Barbara St., Pasadena, CA 91101, USA.
 \item Cahill Center for Astrophysics, California Institute of Technology, Pasadena, CA 91125, USA.
 \item Department of Astronomy, University of Texas, Austin, TX 78712, USA.
 \item The Oskar Klein Centre, Department of Astronomy, Stockholm University, AlbaNova, 10691 Stockholm, Sweden.
 \item Physics Division, Lawrence Berkeley National Laboratory, Berkeley, CA 94720, USA.    
 \item Department of Astronomy, University of California, Berkeley, CA 94720-3411, USA.

\end{affiliations}
\begin{abstract}
\newpage
The explosive fate of massive stripped Wolf-Rayet (W-R) stars\cite{crowther07} is a key open question in stellar physics. An appealing option is that hydrogen-deficient W-R stars are the progenitors of some H-poor supernova (SN) explosions of Types IIb, Ib, and Ic\cite{filippenko97}. A blue object, having
luminosity and colors consistent with those of some W-R stars, has been recently identified at the location of a SN~Ib in pre-explosion
images\cite{cao13} but has not yet been conclusively determined to have been the progenitor. Similar previous works have so far only resulted in
nondetections\cite{eldridge13}. Comparison of early photometric observations of Type Ic supernovae with theoretical models suggests that the progenitor
stars had radii $<10^{12}$\,cm, as expected for some W-R stars\cite{corsi12}. However, the hallmark signature of W-R stars, their emission-line
spectra, cannot be probed by such studies.  
Here, we report the detection of strong emission lines in an early-time spectrum of SN 2013cu (iPTF13ast; Type IIb) obtained $\sim15.5$\,hr after explosion (``flash spectroscopy''). We identify W-R-like wind signatures suggesting a progenitor of the WN(h) subclass. The extent of this dense wind may indicate increased mass loss from the progenitor shortly prior to its explosion, consistent with recent theoretical predictions\cite{shiode13}. 
\end{abstract}

W-R stars are massive stars stripped of their outer H-rich envelope. These stars blow strong H-poor winds. The inner part of the wind engulfing the star is dense and optically thick, and efficiently absorbs the ionizing continuum from the hot stellar surface. Farther from the star, the density drops and the wind becomes optically thin in the continuum, leading to a rich emission spectrum of recombination lines. Detailed models of such spectra can be calculated\cite{crowther07, hamann06} and depend essentially on only three parameters: the effective temperature $T_{\rm eff}$ in the line-forming region, a normalized radius $R_{t}$ (a combination of the stellar radius, luminosity, mass-loss rate, and wind terminal velocity\cite{hamann06}), and the chemical composition $Z$ of the wind (assumed to be uniform, spherical, and of a constant mass-loss rate). The composition of the wind determines W-R spectral classes; stars with dominant He and N lines belong to the WN class (with those also showing traces of H usually denoted as WNh), stars with strong carbon lines belong to the WC class, and rare (and possibly hotter) stars with oxygen-rich spectra reside in the WO class\cite{crowther07}.

Shortly after a W-R star explodes as a SN, the outer parts of the wind (extending in some cases\cite{crowther07} to $>10^{13}$\,cm) that have not yet been swept up by the expanding SN ejecta will emit strong recombination lines in response to ionizing flux released by the explosion shock breakout from the stellar surface. We estimate a factor of order $10^{2}$--$10^{4}$ increase in ionizing luminosity, assuming an initial absolute magnitude range of $-2.5 < M < -10$\,mag for the exploding W-R star\cite{crowther07} and a typical early-time luminosity of $M = -12.5$\,mag for the resulting SN\cite{corsi12,cao13,dessart11}. The effective temperature $T_{\rm eff}$ will also change, being very high ($>10^5$\,K) shortly after explosion\cite{ofek10} and decreasing with time as the shocked SN ejecta cool. The radiation illuminating the surviving W-R wind will thus effectively scan through the range of temperatures in W-R star line-forming regions. Since the wind parameters (composition, mass-loss rate, terminal velocity) do not change, the measured line spectrum observed shortly after explosion should be similar to that of a W-R star having the spectral class of the exploding progenitor (as the spectral classes reflect mainly the wind composition). The high wind densities around W-R stars (with electron densities $n_{\rm e} = 10^{11}$--$10^{12}$\,cm$^{-3}$) imply short recombination times\cite{osterbrock06}, 
$t_{\rm rec} \approx 3.9\times10^{12}\,(n_{\rm e}/{\rm cm}^{-3})^{-1}(T/10^4\,{\rm K})^{0.85}$\,s, typically a few minutes for W-R densities and temperatures, so the emitted spectrum will promptly react to the rapidly evolving SN radiation field.

We obtained rapid spectroscopic observations of the recent Type IIb SN 2013cu (iPTF13ast) shortly after shock breakout (flash spectroscopy; see Methods). This event was first detected by the iPTF survey\cite{law09} on 2013 May 3.18 (UTC dates are used throughout this paper), photometrically confirmed 5.8\,hr later, and promptly identified by an on-duty astronomer who triggered rapid follow-up observations\cite{gal-yam11}, including an optical spectrum obtained just 4\,hr later. Analysis of the early-time light curve of this SN (Extended Data Figure 1) suggests it exploded on May 2.93, implying that the first iPTF detection and the first spectrum correspond to only 5.7\,hr and 15.5\,hr after explosion, respectively. A full description of the SN and its evolution will be reported in a forthcoming publication\cite{gal-yam13}. We note that this event was independently discovered by the MASTER survey on May 5.3 ($\sim2.3$ days after explosion), and it was assigned the name SN 2013cu following spectroscopic confirmation\cite{denisenko13}. 

Our first spectrum of SN 2013cu reveals a continuum and emission lines that bear a striking resemblance to spectra of W-R stars (Fig. 1a). According to accepted W-R terminology\cite{crowther07} the spectrum is classified as a WN6h (Fig. 1a, bottom); the relative strength of nitrogen to carbon lines precludes a WC classification, and the absence of any high-excitation oxygen lines is inconsistent with a WO star. 
The stronger lines (H$\alpha$, H$\beta$, N~IV $\lambda$7115, and He~II $\lambda$5411) exhibit a complex profile (Fig. 2) consisting of a relatively broad base ($\sim2500$\,km\,s$^{-1}$ full width at zero intensity; FWZI) on which prominent narrow, unresolved lines (FWZI $\approx3$\AA; velocity dispersion $<150$\,km\,s$^{-1}$) are superimposed. This is consistent with predictions for W-R pre-SN wind velocities\cite{groh13,nied02}, though we cannot exclude the possibility that at least some of the observed line broadening is produced by
electron scattering rather than genuine velocity dispersion. To the best of our knowledge, no similar spectra of a stripped (H-poor) SN have been acquired previously. W-R-like spectroscopic features have been observed in spectra of some H-rich (nonstripped) SNe obtained at substantially later epochs, and their typically much broader lines were interpreted as emerging from interaction with circumstellar material (CSM)\cite{leonard00}. We further discuss these previous observations in Extended Data Figure~3. 

We analyze our very early spectrum using the PoWR grid of W-R spectral models\cite{hamann06} made available via a web interface at http://www.astro.physik.uni-potsdam.de/$\sim$wrh/PoWR/powrgrid1.html. We find an excellent fit with WN(h) models calculated assuming an H fraction of $20\%$ (by mass) and temperature around $50,000$\,K. This temperature is consistent with the lower limit obtained 
from early {\it Swift} UV photometry (Extended Data Fig. 1). The essentially perfect match of the observed and modeled continuum shapes indicates that dust reddening must be negligible; any pre-existing dust must have been destroyed by the SN explosion flash (see Methods).
We note that among the large catalog of Galactic WN stars specifically modeled in this manner\cite{hamann06}, no stars drive winds which require more than $56\%$ H by mass. Presumably, custom spectral fits\cite{hamann06} could be calculated and used to more accurately determine the physical parameters of the detected W-R wind. Assuming the spectrum was obtained 15.5\,hr after explosion and a standard ejecta velocity of $10^4$\,km\,s$^{-1}$, the narrow-line-emitting material must be located at radii above $\sim5\times10^{13}$\,cm in order not to have been swept up and accelerated by the expanding ejecta. This lower limit is consistent with the extent of some W-R winds, where the line-formation region extends out to several (5--10) times\cite{hillier87} the hydrostatic radius. Recent pre-SN W-R models\cite{groh13} suggest hydrostatic radii of 10--20 solar radii for WN SN progenitors, consistent with line-formation regions extending to several hundred solar radii\cite{crowther07} or $>10^{13}$\,cm.

We further constrain the physical location of the wind using the following calculation. We measure the H$\alpha$ line flux from our spectrum (calibrated to our host-subtracted photometry) and find $F_{\rm H\alpha}=3.4\times10^{-15}$\,erg\,s$^{-1}$\,cm$^{-2}$. We translate this to line luminosity using
a luminosity distance to the host galaxy UGC 9379 of $d=108$\,Mpc, calculated for a flat $\Lambda$CDM cosmology with H$_{0}=73$\,km\,s$^{-1}$\,Mpc$^{-1}$, $\Omega_{\rm m}=0.27$, and a redshift $z=0.025734$, obtained from the NASA Extragalactic Database (NED), as well as negligible extinction, finding  $L_{\rm H\alpha}=4.8\times10^{39}$\,erg\,s$^{-1}$. 
We can then estimate\cite{ofek13} the pre-explosion H mass-loss rate $\dot{M}=0.01 \times (L_{\rm H\alpha}/2\times10^{39}\,{\rm erg\,s}^{-1})^{1/2}\,(v_w/500\,{\rm km\,s}^{-1})\,(r/10^{15}\,{\rm cm})^{1/2}$\,M$_{\odot}$\,yr$^{-1}$ assuming the lines are formed at a radius $r$ via recombination, an emitting shell whose width is similar to its radius, spherical symmetry, and a wind profile with density falling as $r^{-2}$. We assume a wind velocity $v_w=2500$\,km\,s$^{-1}$, consistent with our spectra and as expected for W-R stars, but our results are not sensitive to this value and remain essentially unchanged for $100<v_w<2500$\,km\,s$^{-1}$. 

We check for self consistency by calculating the implied electron density, and hence the Thomson optical depth $\tau=0.3\,(\dot{M}/0.01\,{\rm M}_{\odot}\,{\rm yr}^{-1})\,(v_w/500\,{\rm km\,s}^{-1})^{-1}\,(r/10^{15}\,{\rm cm})^{-1}$ at this same radius, and require it to be lower than $\tau=1$ for the lines to escape. We find that this self-consistency requirement places a lower limit on the line-formation region $r>2\times10^{14}$\,cm, with substantial mass-loss
rates $\dot{M}>0.03$\,M$_{\odot}$\,yr$^{-1}$.  Interpreting the disappearance of essentially all emission lines from our day 6 spectrum (Fig. 1b) as evidence that the wind was swept up by the expanding ejecta (moving at $10^4$\,km\,s$^{-1}$), the radius of the line-emitting region must be $r<5.2\times10^{14}$\,cm, fully consistent with our estimates.   

We can then calculate the total H mass by integrating over $r$: $M_{\rm tot}= 0.006\,(\dot{M}/0.01\,{\rm M}_{\odot}\,{\rm yr}^{-1})\,\times (v_w/500\,{\rm km\,s}^{-1})^{-1}\,(r/10^{15}\,{\rm cm})^{-1}$\,M$_{\odot}$, indicating a range of $0.0008\,{\rm M}_{\odot}<M_{\rm tot}<0.0035\,{\rm M}_{\odot}$ for the range of permitted H masses. Assuming the typical H abundances for WN(h) stars ($\sim20\%$), the total wind mass (dominated by He) can be estimated to be several times larger than these numbers. Detailed simulations\cite{chugai09} show that as little as $0.1$\,M$_{\odot}$ of He-dominated CSM would result in strong spectroscopic interaction signatures (that we do not observe), consistent with our derived total masses.

We conclude that we have directly detected a W-R-like wind from the SN progenitor with a WN(h) spectral class, indicating a low H mass fraction. 
Assuming the wind composition we measure represents the surface composition of the progenitor star, our observations indicate that 
some members of the spectroscopic WNh W-R class explode after having lost most of the hydrogen envelope, exposing the CNO-processed, N-rich He layer below. Analysis of photometric and spectroscopic follow-up observations\cite{gal-yam13} confirms that the explosion was indeed a SN of Type IIb (Fig. 1c), as expected if the progenitor was a massive star that lost all but $\sim0.1$\,M$_{\odot}$ of its H envelope\cite{hachinger12}.

Our observations have interesting implications. 
First, we note that the derived value of the mass-loss rate and emission-line-region size are
quite extreme compared with known W-R observations and radiatively driven models\cite{vink11}, including models with clumpy, inflated atmospheres\cite{graefener12},
suggesting that the mass-loss rate from the progenitor star may have increased shortly
(of order one year for the assumed velocities) prior to its explosion. Interestingly, such pre-SN activity may be explained by recent wave-driven models\cite{shiode13}, or perhaps more extreme envelope inflation\cite{graefener12} is indicated. These data can thus provide a key diagnostic of the final stages of nuclear core burning in
massive stars, currently poorly understood, with potential deep implications into the explosion mechanism itself. In any case,  
the star probably exploded inside a thick wind, and the explosion shock may have broken out from the opaque inner wind rather than from the hydrostatic surface of the star\cite{ofek10}. 

Our finding is in general accord with some previous work on SN~IIb progenitors. Direct imaging of the progenitor of SN 2008ax\cite{crockett08} is consistent with a WNh progenitor. Furthermore, increased mass loss during the final year prior to explosion may inflate the apparent photospheric radius of the pre-SN star, making stars with compact cores appear to have extended (low-mass) envelopes\cite{graefener12}, possibly reconciling the conflicting findings about
the progenitor of the Type IIb SN 2011dh\cite{arcavi11,bersten12,vandyk13,ergon13}. Regardless of the exact mechanism, our observations suggest that substantial W-R-like winds predate at least some SNe~IIb. A strong metallicity dependence of this process may explain the trend in the SN~IIb/SN~Ib ratios with host-galaxy metallicity\cite{arcavi10}. Future studies of numerous additional supernova progenitors via their spectroscopic wind signatures (see Methods) would provide powerful constraints on the final stages of massive-star evolution. 

\noindent \underline{Methods Summary}
Photometry: $r$-band observations were obtained by the iPTF survey telescope. Photometry
is measured using our custom pipeline performing point-spread-function (PSF) photometry on iPTF images after removing a reference image 
constructed from pre-explosion data using image subtraction.  {\it Swift} UV absolute AB magnitudes (Extended Data Fig. 1) are measured using standard pipeline reduction and are corrected for host-galaxy contamination using late-time {\it Swift} images. Spectroscopy: our earliest (15.5 hr) and latest (69 days) spectra were obtained using the DEIMOS spectrograph mounted on the Keck-II 10\,m telescope using the 600\,line\,mm$^{-1}$ grating and an exposure time of 600\,s. Additional spectra were obtained
using ALFOSC mounted on the 2.56\,m NOT telescope, LRS mounted on the 10.4\,m HET telescope (day 3), and LRIS mounted on the Keck-I 10\,m telescope (day 6). All spectra were reduced using standard pipelines and are digitally available on WISeREP. The method of flash spectroscopy is described in detail in the extended methods section.

\include{PTF13ast_EndNotes_alex_v2}

\newpage

\begin{figure}[h!]
\includegraphics[width=16cm]{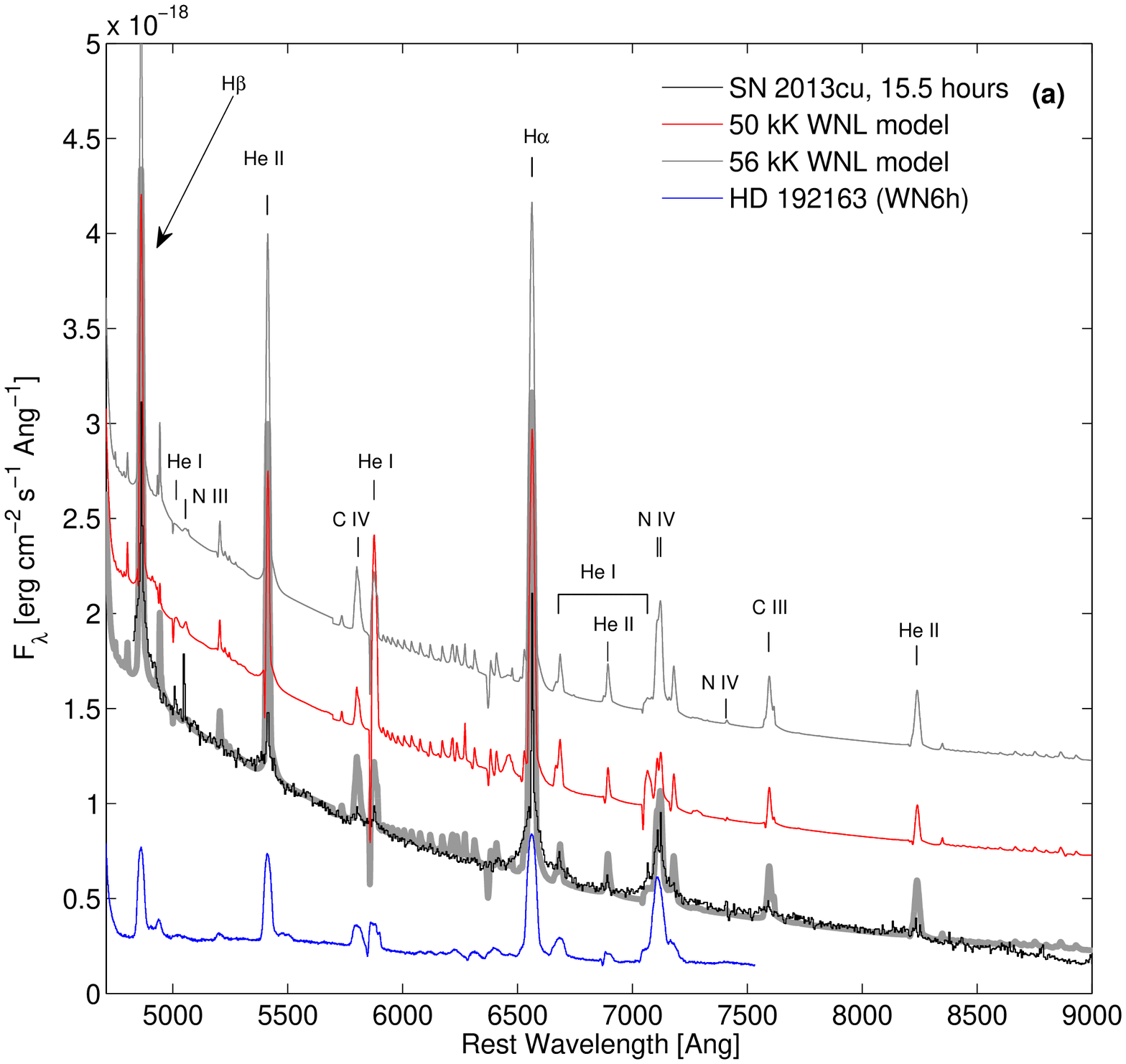}
\caption*{Figure 1: Spectroscopy of the Type IIb SN 2013cu reveals transient W-R-like features. {\bf (a)} The early spectrum of SN 2013cu (black) is compared with WNL models (red and gray curves)\cite{hamann06} showing remarkable similarity in line features (strong He, N and Balmer lines indicate a WN6h classification) and in the continuum. {\bf (b)} Emission-line evolution during the first week. The first spectrum (red) is compared with later spectra. By day 3 (blue+cyan) the initially strong W-R features disappear, while the H$\alpha$ line remains constant at $3.4\times10^{-15}$\,erg\,s$^{-1}$\,cm$^{-2}$. The spectrum on day 6 (magenta) is almost featureless, except for weak H$\alpha$ emission (inset) with an intensity $<0.1$ that of day 3, likely suggesting that the line-forming region has been cleared by the expanding ejecta. {\bf (c)}: SN 2013cu is a SN~IIb.  A spectrum of SN 2013cu 69 days after explosion (black) is compared with the prototypical Type IIb SNe 1993J (+60 days; red) and 2011dh (+43 days; green), and with the typical nonstripped Type II-P SN 2004et (45 days; blue). SNe 2013cu, 2011dh, and 1993J exhibit strong He~I absorption at 5876\,\AA, 7065\,\AA, and 7281\,\AA\ (marked with black vertical lines), which are not detected in SNe~II-P. See methods section for additional details. 
}
\end{figure}

\newpage
\begin{figure}[h!]
\includegraphics[width=16cm]{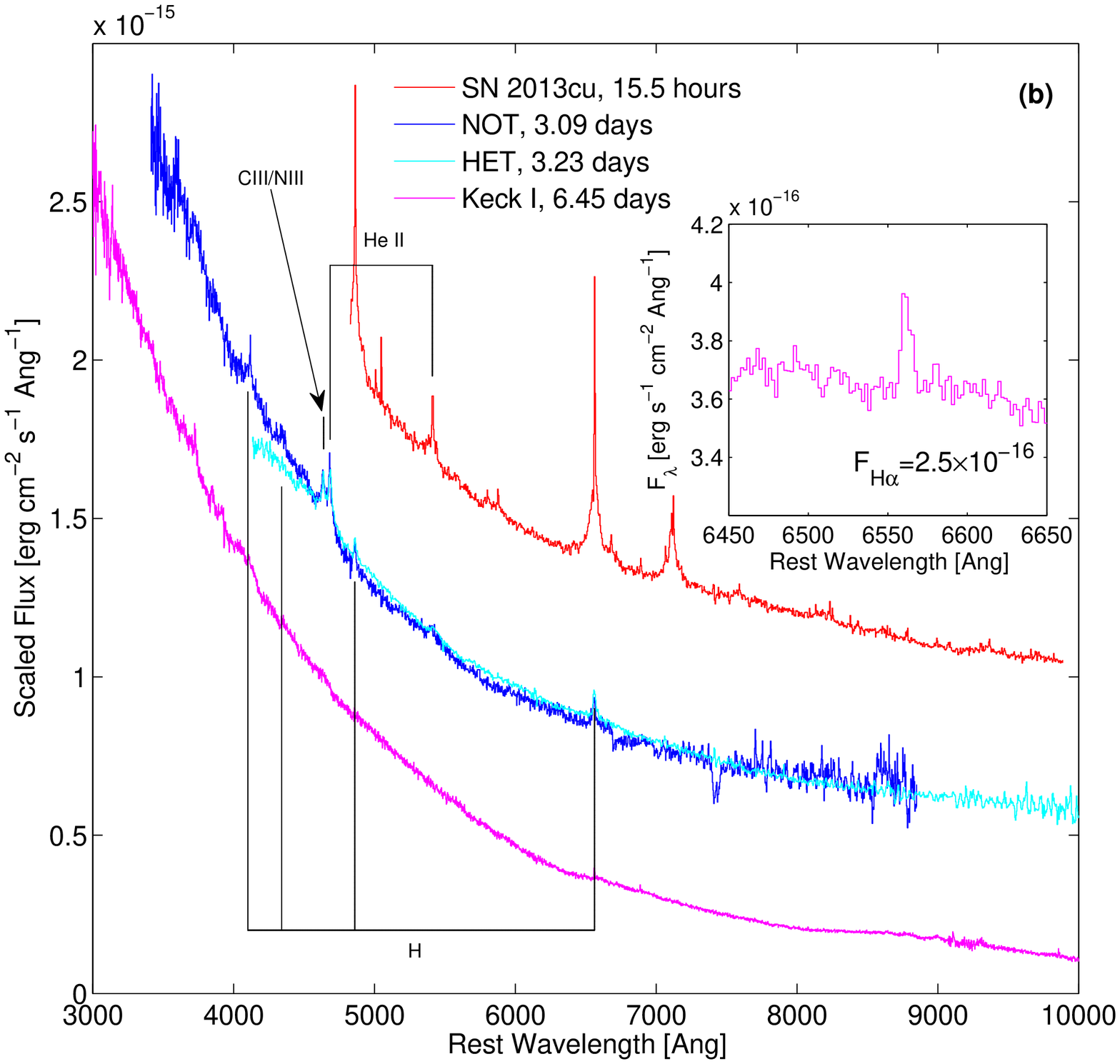}
\caption*{ 
}
\end{figure}

\newpage
\begin{figure}[h!]
\includegraphics[width=16cm]{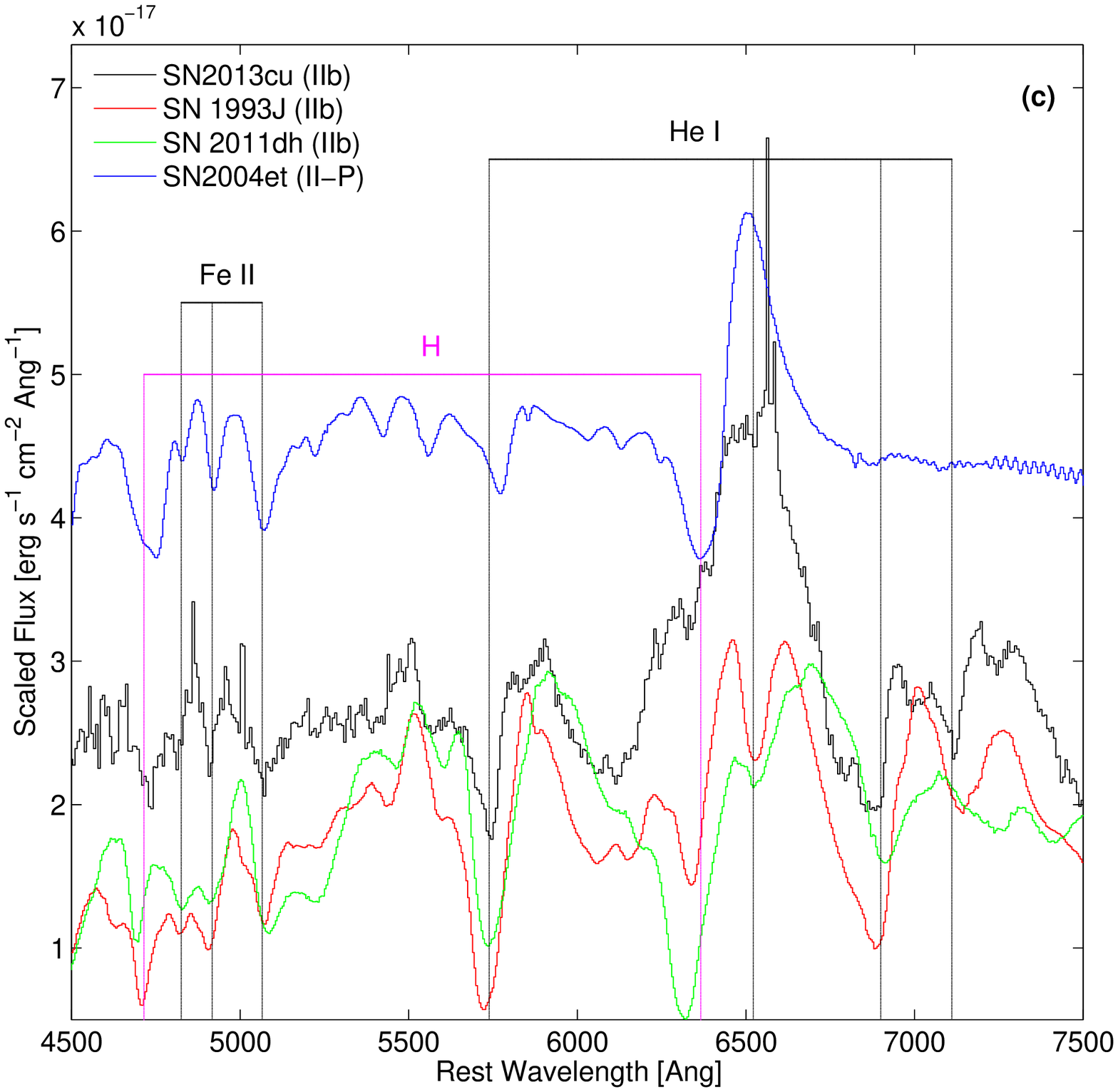}
\caption*{
}
\end{figure}

\newpage

\begin{figure}[h!]
\includegraphics[width=16cm]{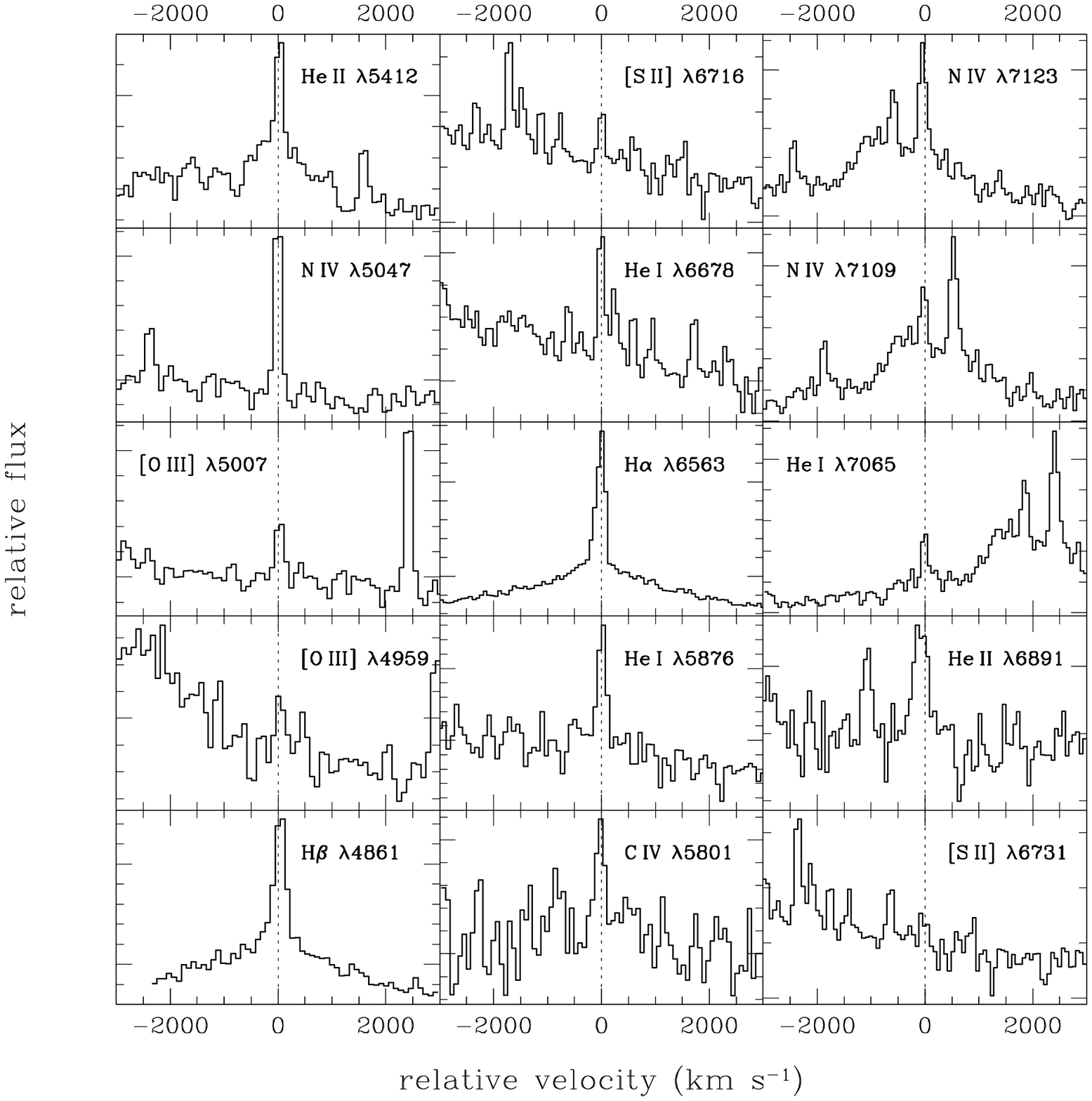}
\caption*{Figure 2: Emission-line velocity structure at 15.5\,hr. The strongest lines (H$\alpha$ and H$\beta$, He~II $\lambda$5411, and the N~IV $\lambda$7115 complex) show broad wings extending out to $\sim2500$\,km\,s$^{-1}$. Other weaker lines are narrow and unresolved.  
}
\end{figure}

\newpage

\begin{figure}[h!]
\includegraphics{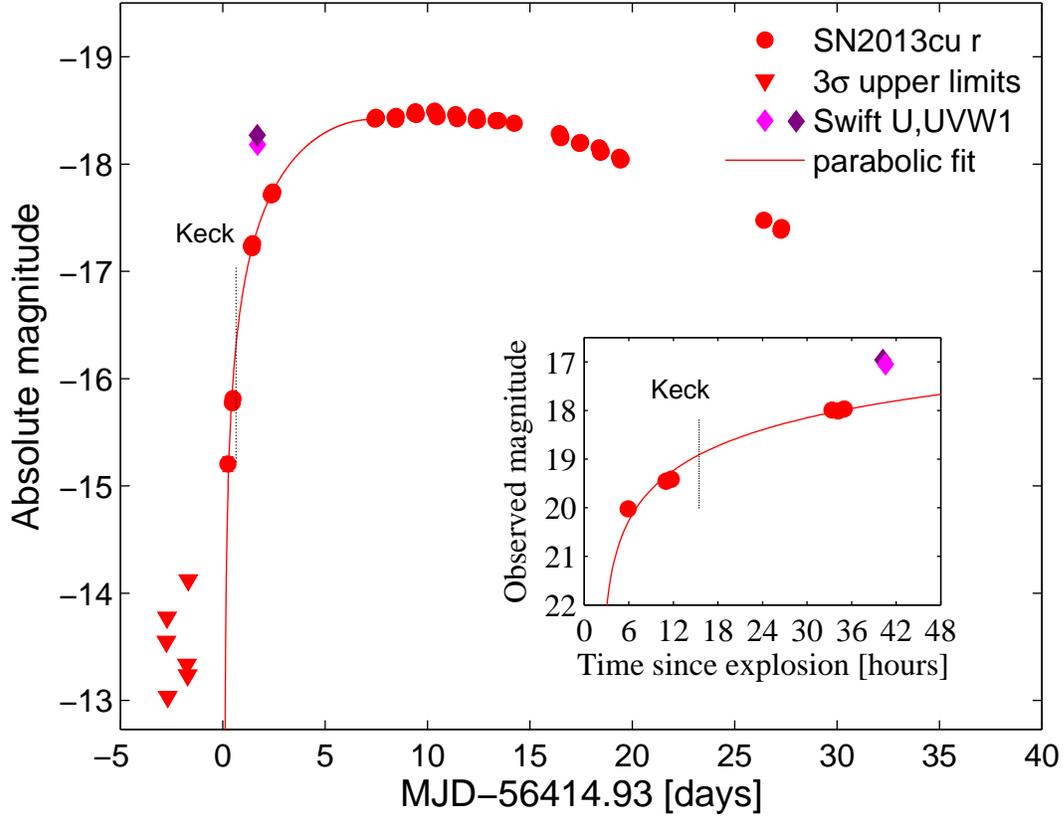}
\caption*{Extended Data Figure 1: The $r$-band light curve of SN 2013cu. A parabolic model of the flux vs. time (red solid curve) describes the pre-peak data (1$\sigma$ s.d. error bars) very well. Backward extrapolation indicates an explosion date of UTC 2013 May $2.93 \pm 0.11$ (MJD = 56414.93; 5.7\,hr before the first 
iPTF detection; see inset); we estimate the uncertainty from the scatter generated by modifying the subset of points used in the fit. Our first Keck spectrum was obtained about 15.5\,hr after explosion (vertical dotted line). Early {\it Swift} UV photometry (diamonds) places a lower limit of $T=$ 25,000\,K on the blackbody temperature measured 40\,hr
after explosion.}
\end{figure}

\newpage

\begin{figure}[h!]
\includegraphics[width=12cm]{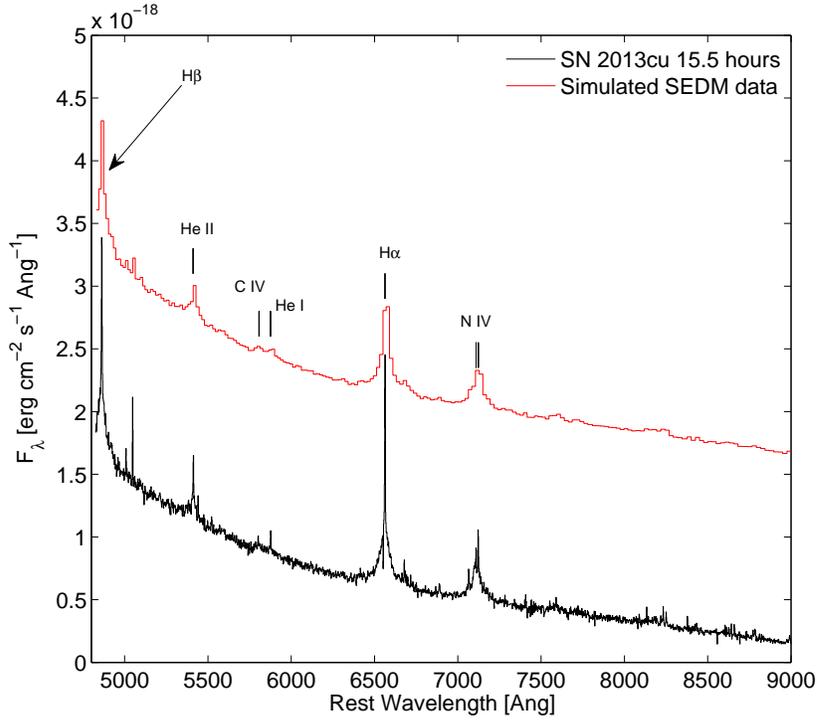}
\caption*{Extended Data Figure 2: Flash spectroscopy --- rapid spectroscopic observations of SNe during or shortly after shock breakout. This simulated SEDM spectrum (red) created by downgrading the observed Keck spectrum (black; resolution $R=2000$) to the coarse SEDM resolution ($R=100$) shows that the strong W-R lines (in this case the marked H, He, and N lines) are still easily detectable and allow us to determine the W-R spectroscopic class. The SEDM is an IFU low-resolution spectrograph designed for
robotic response to transient events, to be mounted on the Palomar 60-inch telescope
almost continuously. Responding to real-time triggers from the iPTF survey operating on the same mountain, this instrument should be able to obtain
low-resolution spectra within $\sim1$\,hr of object detection. Operating on a smaller telescope than Keck, SEDM data
of similar quality to the simulated spectrum will require a relatively long integration. However, SEDM will be able to 
observe objects with much reduced latency, benefiting from stronger line intensities expected owing to stronger shock-breakout flash luminosity processed by a denser wind close to the progenitor star, potentially compensating for its reduced absolute sensitivity.
}
\end{figure}

\newpage

\begin{figure}[h!]
\includegraphics[width=14cm]{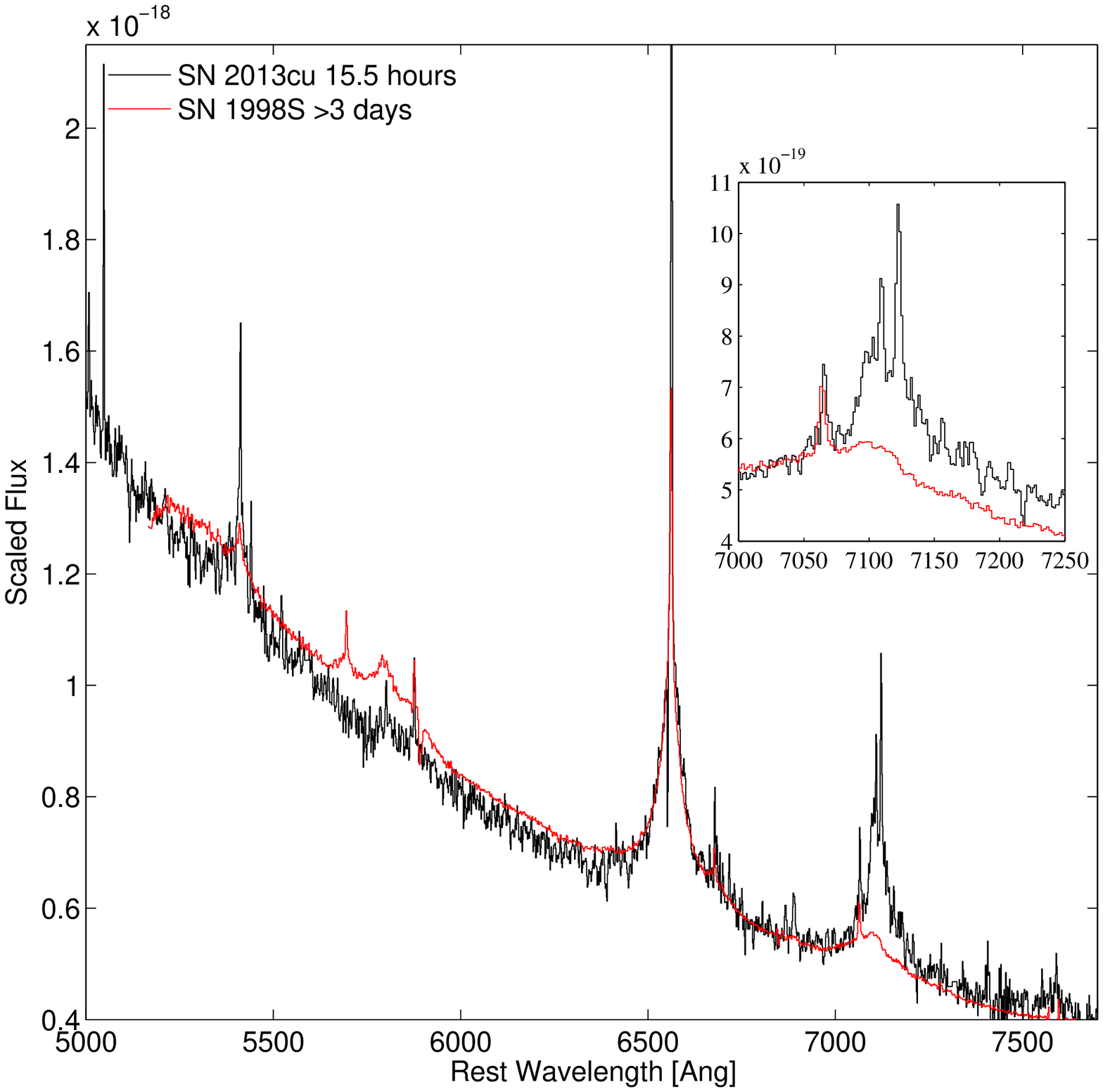}
\caption*{Extended Data Figure 3: Comparison with early ``W-R'' spectra of SN 1998S\cite{leonard00}. W-R-like features similar to those we observed 
were previously noted in two
cases, SN 1983K$^{40,41}$ and SN 1998S, and persisted for many days after explosion. 
The spectra of SN 1983K, classified as a SN~II-P, are unfortunately not available for comparison. Spectra
of SN 1998S (Type IIn) are shown here. 
The spectra have a similar blue continuum slope and a similar H$\alpha$ profile. The He~II $\lambda$5411 and the N~IV $\lambda$7115 complex are weaker in SN 1998S, and the strong lines of N and C (5806\,\AA) are
broad, consistent with an origin in shocked CSM\cite{leonard00} rather than in an undisturbed W-R wind.  The inset shows a close-up view of the
strong N~IV $\lambda$7115 complex. N emission from SN 1998S is weak compared with He~I $\lambda$7065 and shows a smooth, broad profile, while SN 2013cu
exhibits a broad base (2500\,km\,s$^{-1}$ FWHM) as well as strong and narrow (unresolved) N~IV lines. These observations are consistent with an origin of the N emission of SN 1998S in shocked (and perhaps N-rich) CSM\cite{leonard00}, while the narrow He~I lines may come from a more distant, photoionized wind. In SN 2013cu even the narrow N lines are much stronger than He, indicating a wind with W-R-like composition at all radii.}
\end{figure}

\newpage

\noindent \underline{Methods}

Photometry: $r$-band images were obtained by the iPTF survey camera mounted on the Palomar 48-inch Schmidt telescope$^{11,31}$. Photometry
is measured using our custom pipeline performing PSF photometry on iPTF images after removing a reference image 
constructed from pre-explosion data using image subtraction.  {\it Swift} UV absolute AB magnitudes (Extended Data Fig. 1) are measured using standard pipeline reduction$^{32}$ and are corrected for host-galaxy contamination using late-time {\it Swift} images. 

Spectroscopy: our earliest (15.5 hr) and latest (69 days) spectra were obtained using the DEep Imaging Multi-Object Spectrograph (DEIMOS$^{33}$) spectrograph mounted on the Keck-II 10\,m telescope using the 600\,line\,mm$^{-1}$ grating and an exposure time of 600\,s. Additional spectra were obtained
using ALFOSC mounted on the 2.56\,m NOT telescope, LRS mounted on the 10.4\,m HET telescope (day 3), and the Low Resolution Imaging Spectrometer (LRIS$^{34}$) mounted on the Keck-I 10\,m telescope (day 6). All spectra were reduced using standard pipelines and are digitally available on WISeREP$^{35}$. 

Spectroscopic observations are presented in Fig. 1. {\bf (a)} The early spectrum of SN 2013cu reveals W-R wind features. The spectrum (black) is compared with WNL models (red and gray curves, offset vertically for clarity)\cite{hamann06} showing remarkable similarity, both in line features (major species marked; strong He and N lines accompanied by Balmer lines indicate a WN6h classification) and in the continuum shape (demonstrated by overplotting the 56\,kK model on the spectrum). The similarity in continuum shape to hot model spectra limits any dust reddening to be minimal, indicating that any pre-existing circumstellar dust must have been destroyed; compare with the observed spectrum of the WN6(h) star HD 192163 (blue). Consistent with this conclusion, we detect no trace of Na~D absorption lines. {\bf (b)} Emission-line evolution during the first week. We replot the first spectrum (red), compared with spectra obtained $\sim3$ days after explosion (using ALFOSC, blue; and LRS, cyan) and $\sim6$ days after explosion (magenta; see Ref.\cite{gal-yam13} for additional details). Spectra were scaled and offset for clarity with respect to the high-quality Keck/LRIS spectrum. Continuum shapes are identical (within calibration uncertainties) and consistent with the Rayleigh-Jeans tail of a hot Planck curve, indicating that the blackbody peak remains in the UV at 6 days past explosion. By day 3 the initially strong W-R features disappear, with the exception of weaker He~II lines and the C~III/N~III complex around 4640\,\AA\ (marked), while the H$\alpha$ line remains constant at $3.4\times10^{-15}$\,erg\,s$^{-1}$\,cm$^{-2}$ (with decreased equivalent width due to higher continuum level). The high signal-to-noise ratio spectrum on day 6 is almost featureless, except for weak H$\alpha$ emission (inset) with an intensity $<0.1$ that of day 3, likely suggesting that the line-forming region has been cleared by the expanding ejecta. {\bf (c)}: SN 2013cu is a SN~IIb. A spectrum of SN 2013cu 69 days after explosion (black) is compared with the prototypical Type IIb SNe 1993J at age 60 days (red) and 2011dh at age 43 days (green), and with the typical nonstripped Type II-P SN 2004et at age 45 days (blue). To allow for slight age differences and expansion-velocity variations, we align all spectra in wavelength using the weak Fe~II lines at 4924\,\AA, 5018\,\AA, and 5169\,\AA\ (marked with black vertical lines), as they are good tracers of the photosphere. SNe 2013cu, 2011dh, and 1993J exhibit strong He~I absorption at 5876\,\AA, 7065\,\AA, and 7281\,\AA\ (marked with black vertical lines), while the weaker 6678\,\AA\ absorption is similar in SN 2013cu and SN 2011dh. These He~I lines are not detected in the SN~II-P spectrum at all. On the other hand, SN 2013cu shows weaker Balmer absorption (marked with magenta vertical lines), and the H$\alpha$ absorption is not clearly defined. Both the spectral similarity to SNe 1993J and 2011dh and the strong He~I lines compared to H indicate that SN 2013cu is spectroscopically a SN~IIb.  

Flash spectroscopy: we define as ``flash spectroscopy'' a set of spectroscopic data obtained shortly enough after a supernova explosion so that the observed spectrum is dominated by features related directly to the effects of the shock-breakout flash. In particular, flash-ionized CSM recombines and forms strong emission lines, revealing, for example, the elemental abundance and thus the W-R class of a SN progenitor. 
In addition, emission-line spectra provide a handle on the early temperature evolution, which is difficult to measure even using {\it Swift} UV photometry because the blackbody peak is initially too far into the UV.  This study provides strong motivation for future investigations using dedicated rapid-response spectrographs such as FLOYDS$^{3,36}$ and SEDM$^{37}$ responding to real-time triggers from high-cadence wide-field surveys\cite{gal-yam11}. 

While W-R SN progenitor stars are difficult to study using pre-explosion imaging (owing to both intrinsic low luminosity in the optical/infrared bands\cite{yoon12, groh13} and the possible confusing effect of a bright O/B companion\cite{cao13}), we demonstrate that they are amenable to study using the flash spectroscopy method. W-R stars belonging to the WNh class may have the most extensive winds\cite{crowther07,groh13}. Application of this
method to WC/WO stars may require flash spectroscopy at even earlier epochs ($\sim1$\,hr after explosion), before the SN ejecta sweep up the high-density wind. Reducing the latency between SN explosion and spectroscopy by an order of magnitude compared to our observations of SN 2013cu is possible
using recently commissioned instrumentation (Extended Data Fig. 2). 

Unlike studies of SN progenitors through pre-explosion imaging, the flash spectroscopy method can be applied to relatively distant objects (SN 2013cu is located 108\,Mpc away, well beyond the 20\,Mpc distance typical for pre-explosion studies), and to events in galaxies having no pre-explosion high-quality imaging, such as a large population of little-studied dwarf galaxies. Based on local SN rate measurements$^{38}$, $\sim300$ events explode within 100\,Mpc every year and can be potentially studied in this manner. The method thus allows routine spectroscopic studies of SN progenitors, previously only possible
by extreme serendipity (e.g., for the progenitor of SN 1987A$^{39}$).
Within a few years, the flash spectroscopy method can be used to chart wind signatures from numerous SN progenitors, and in particular, the W-R progenitor population of stripped SNe may be studied systematically. We thus expect that this method will be broadly applied in the coming years.
\begin{itemize}
\item[]
{[31] {Rau}, A., {Kulkarni}, S.~R., {Law}, N.~M., {Bloom}, J.~S., {Ciardi}, D., {Djorgovski}, G.~S., et al.}
{Exploring the Optical Transient Sky with the Palomar Transient Factory}
\newblock {\it PASP} {\bf 121}, 1334-1351 (2009)
\item[]
{[32] {Poole}, T.~S., {Breeveld}, A.~A., {Page}, M.~J., {Landsman}, W., {Holland}, S.~T., {Roming}, P., et al.}
{Photometric calibration of the Swift ultraviolet/optical telescope}
\newblock {\it MNRAS} {\bf 383}, 627-645 (2008)
\item[]
{[33] {Faber}, S.~M., {Phillips}, A.~C., {Kibrick}, R.~I., {Alcott}, B., {Allen}, S.~L., {Burrous}, J., et al.}
{The DEIMOS spectrograph for the Keck II Telescope: integration and testing}
\newblock {\it SPIE} {\bf 4841}, 1657-1669 (2003)
\item[]
{[34] {Oke}, J.~B., {Cohen}, J.~G., {Carr}, M., {Cromer}, J., {Dingizian}, A., {Harris}, F.~H., et al.}
{The Keck Low-Resolution Imaging Spectrometer}
\newblock {\it PASP} {\bf 107}, 375 (1995)
\item[]
{[35] {Yaron}, O., \& {Gal-Yam}, A.}
{WISeREP - An Interactive Supernova Data Repository.}
\newblock {\it PASP} {\bf 124}, 668-681 (2012)
\item[]
{[36] {Ben-Ami}, S., {Gal-Yam}, A., {Filippenko}, A.~V., {Mazzali}, P.~A., {Modjaz}, M., {Yaron}, O., et al.}
{Discovery and Early Multi-wavelength Measurements of the Energetic Type Ic Supernova PTF12gzk: A Massive-star Explosion in a Dwarf Host Galaxy.}
\newblock {\it ApJ} {\bf 760}, L33 (2012) 
\item[]
{[37] {Ben-Ami}, S., {Konidaris}, N., {Quimby}, R., {Davis}, J.~T., {Ngeow}, C.~C., {Ritter}, A., et al.}
{The SED Machine: a dedicated transient IFU spectrograph.}
\newblock {\it Proc. SPIE} {\bf 8446}, 844686 (2012) 
\item[]
{[38] {Li}, W., {Leaman}, J., {Chornock}, R., {Filippenko}, A.~V., {Poznanski}, D., {Ganeshalingam}, M., et al.}
{Nearby supernova rates from the Lick Observatory Supernova Search - II. The observed luminosity functions and fractions of supernovae in a complete sample.}
\newblock {MNRAS} {\bf 412}, 1441-1472 (2011) 
\item[]
{[39] {Walborn}, N.~R., {Prevot}, M.~L., {Prevot}, L., {Wamsteker}, W., {Gonzalez}, R., {Gilmozzi}, R. et al..}
{The spectrograms of Sanduleak -69.202 deg, precursor to supernova 1987A in the Large Magellanic Cloud}
\newblock {A\&A} {\bf 219}, 229-236 (1989) 
\item[]
{[40] {Niemela}, V.~S., {Ruiz}, M.~T. and {Phillips}, M.~M.}
{The supernova 1983k in NGC 4699 - Clues to the nature of Type II progenitors}
\newblock {ApJ} {\bf 289}, 52-57 (1985) 
\item[]
{[41] {Phillips}, M.~M., {Hamuy}, M., {Maza}, J., {Ruiz}, M.~T., {Carney}, B.~W. and {Graham}, J.~A.}
{The light curve of the plateau Type II SN 1983K}
\newblock {PASP} {\bf 102}, 299-305 (1990) 

\end{itemize}

\end{document}

%% file: PTF13ast_EndNotes_alex_v2.tex
\newenvironment{addendum}{%
    \setlength{\parindent}{0in}%
    \small%
    \begin{list}{Acknowledgments}{%
        \setlength{\leftmargin}{0in}%
        \setlength{\listparindent}{0in}%
        \setlength{\labelsep}{0em}%
        \setlength{\labelwidth}{0in}%
        \setlength{\itemsep}{12pt}%
        \let\makelabel\addendumlabel}
    }
    {\end{list}\normalsize}

\newcommand*{\addendumlabel}[1]{\textbf{#1}\hspace{1em}}

\begin{addendum}
\appendix
\item[Supplementary Information] 
is linked to the online version of this paper at www./nature.com/nature.
\item[Acknowledgments]
This research was supported by the I-CORE Program ``The Quantum Universe'' of the Planning and Budgeting Committee and The Israel Science Foundation.
A.G. acknowledges support by grants from the ISF, BSF, GIF, Minerva, the FP7/ERC, and a Kimmel Investigator award.
M.M.K. acknowledges generous support from the Hubble and Carnegie-Princeton Fellowships.
E.O.O. acknowledges the Arye Dissentshik career development chair and a grant from the Israeli MOST.
J.C.W. is supported in part by the NSF.
J.M.S. is supported by an NSF Postdoctoral Fellowship.
A.V.F. acknowledges financial support from the TABASGO Foundation, the
Richard \& Rhoda Goldman Fund, the 
Christopher R. Redlich Fund, and the NSF.
The National Energy Research Scientific Computing Center, supported by the Office of Science of the U.S. Department of Energy, provided staff, computational resources, and data storage for this project.
The OKC is funded by the Swedish Research Council. 
K. Clubb, O. Fox, P. Kelly, S. Tang, and B. Sesar are thanked for their help with observations, and J. Groh, P. Crowther, M. Bersten, and E. Nakar for helpful advice.
Some of the data presented herein were obtained at the W. M. Keck
Observatory, which is operated as a scientific partnership among the
California Institute of Technology, the University of California, and
NASA; the observatory was made possible by the generous financial
support of the W. M. Keck Foundation.

\item[Author Contributions]
A.G. initiated the study, conducted analysis, and wrote the manuscript.
I.A. found the SN, triggered rapid follow-up spectroscopy, and contributed to the light-curve analysis, observations, data reduction, and manuscript preparation.
E.O.O. contributed to analysis of early-time data, mass-loss estimates, temperature evolution, and manuscript preparation.
S.B. contributed to data reduction, and to early light-curve and spectroscopic analysis.
S.B.C. reduced {\it Swift} and Palomar 60-inch data, and contributed to spectroscopic reduction and analysis.
M.M.K. provided APO data and contributed to the manuscript preparation.
Y.C. contributed to APO data reduction, early light-curve analysis, and manuscript preparation.
O.Y. contributed to observations and manuscript preparation.
D.T. provided unpublished SN light-curve templates and contributed to photometric analysis.
J.M.S. provided spectroscopic reduction and advice, and contributed to HET spectroscopy.
A.H. provided early Keck spectroscopy.
A.D. contributed to observations and data reduction.
F.T. reduced NOT data.
J.S. provided NOT spectroscopy.
D.P. provided Keck spectroscopy and analysis.
P.V. assisted with observations, spectroscopic analysis, figure preparation, and manuscript writing.
P.E.N. is a PTF builder and contributed to the manuscript.
S.R.K. is a PTF builder.
A.V.F. provided Keck data and edited the manuscript.
J.C.W. provided HET data.
 
\item[Author Information]
Reprints and permissions information is available at www.nature.com/reprints.
Correspondence and requests for materials should be addressed to A. Gal-Yam (e-mail: avishay.gal-yam@weizmann.ac.il).




\end{addendum}

%% file: PTF13ast-revised-alex.bbl
\begin{thebibliography}{}
\newcommand{\aap}{Astron. Astrophys.}
\newcommand{\araa}{Ann. Rev. Astron. Astrophys.}
\newcommand{\apj}{Astrophys. J.}
\newcommand{\aj}{Astron. J.}
\newcommand{\apjl}{Astrophys. J. Letters}
\newcommand{\apjs}{Astrophys. J. Supplements}
\newcommand{\nat}{Nature}
\newcommand{\pasp}{Publications of the Astronomical Society of the Pacific}
\newcommand{\mnras}{Mon. Not. R. Astron. Soc.}
\newcommand{\sci}{Science}
\newcommand{\prd}{Phys.R. D}

\bibitem[1]{crowther07} 
{{Crowther}, P.~A.}
{Physical Properties of Wolf-Rayet Stars.}
\newblock {\it \araa} {\bf 45}, 177-219 (2007) 

\bibitem[2]{filippenko97} 
{{Filippenko}, A.~V.}
{Optical Spectra of Supernovae.}
\newblock {\it \araa} {\bf 35}, 309-355 (1997) 

\bibitem[3]{cao13} 
{{Cao}, Y., {Kasliwal}, M.~M., {Arcavi}, I., {Horesh}, A., {Hancock}, P., {Valenti}, S., et al.}
{Discovery, Rise and Progenitor of a Stripped Envelope Supernova iPTF13bvn.}
\newblock {\it \apjl} {\bf 775}, L7 (2013)

\bibitem[4]{eldridge13} 
{{Eldridge}, J.~J., {Fraser}, M., {Smartt}, S.~J., {Maund}, J.~R., {Crockett}, R.~M.}
{The death of massive stars - II. Observational constraints on the progenitors of type Ibc supernovae.}
\newblock {\it \mnras} {\bf 436}, 774 (2013) 

\bibitem[5]{corsi12} 
{{Corsi}, A., {Ofek}, E.~O., {Gal-Yam}, A., {Frail}, D.~A., {Poznanski}, D., {Mazzali}, et al.}
{Evidence for a Compact Wolf-Rayet Progenitor for the Type Ic Supernova PTF 10vgv.}
\newblock {\it \apj} {\bf 741}, L5 (2012)

\bibitem[6]{shiode13} 
{{Shiode}, J.~H., {Quataert}, E.}
{Setting the Stage for Circumstellar Interaction in Core-Collapse Supernovae II: Wave-Driven Mass Loss in Supernova Progenitors}
\newblock {\it \apj} {\bf arXiv/1308.5978}, Submitted (2013) 

\bibitem[7]{hamann06} 
{{Hamann}, W.-R., {Gr{\"a}fener}, G., {Liermann}, A.}
{The Galactic WN stars. Spectral analyses with line-blanketed model atmospheres versus stellar evolution models with and without rotation.}
\newblock {\it \aap} {\bf 457}, 1015-1031 (2006)

\bibitem[8]{dessart11} 
{{Dessart}, L., {Hillier}, D.~J., {Livne}, E., {Yoon}, S.-C., {Woosley}, S., {Waldman}, R., et al.}
{Core-collapse explosions of Wolf-Rayet stars and the connection to Type IIb/Ib/Ic supernovae.}
\newblock {\it \mnras} {\bf 414}, 2985-3005 (2011)

\bibitem[9]{ofek10}
 {{Ofek}, E.~O., {Rabinak}, I., {Neill}, J.~D., {Arcavi}, I., {Cenko}, S.~B., {Waxman}, E., et al.}
{Supernova PTF 09UJ: A Possible Shock Breakout from a Dense Circumstellar Wind.}
\newblock {\it \apj} {\bf 724}, 1396-1401 (2010)

\bibitem[10]{osterbrock06} 
{{Osterbrock}, D.~E., {Ferland}, G.~J.}
{Astrophysics of Gaseous Nebulae and Active Galactic Nuclei.}
\newblock {Sausalito, CA}, Univ. Science Books (2006)

\bibitem[11]{law09} 
{{Law}, N.~M., {Kulkarni}, S.~R. {Dekany}, R.~G., {Ofek}, E.~O., {Quimby}, R.~M., {Nugent}, P.~E., et al.}
{The Palomar Transient Factory: System Overview, Performance, and First Results.}
\newblock {\it \pasp} {\bf 121}, 1395-1408 (2009) 

\bibitem[12]{gal-yam11}
 {{Gal-Yam}, A., {Kasliwal}, M.~M., {Arcavi}, I., {Green}, Y., {Yaron}, O., {Ben-Ami}, S., et al.}
{Real-time Detection and Rapid Multiwavelength Follow-up Observations of a Highly Subluminous Type II-P Supernova from the Palomar Transient Factory Survey.}
\newblock {\it \apj} {\bf 736}, 159 (2011)

\bibitem[13]{gal-yam13}
 {{Gal-Yam}, A., et al.}
{Discovery of the Type IIb Supernova SN 2013cu During the Shock Breakout.}
\newblock {\it \apj} in prep. (2014)

\bibitem[14]{denisenko13}
 {{Denisenko}, D., {Shurpakov}, S., {Gorbovskoy}, E., {Sarneczky}, K., {Tomasella}, L., {Benetti}, S., et al.}
{SUPERNOVA 2013cu IN UGC 9379 = PSN J14335897+4014207.}
\newblock {CBET} {\bf 3540}, (2013)

\bibitem[15]{groh13} 
{{Groh}, J.~H., {Meynet}, G., {Georgy}, C., {Ekstrom}, S.}
{Fundamental properties of core-collapse Supernova and GRB progenitors: predicting the look of massive stars before death.}
\newblock {\it \aap} {\bf 558}, 131 (2013) 

\bibitem[16]{nied02} 
{{Niedzielski}, A. and {Skorzynski}, W.}
{Kinematical Structure of Wolf-Rayet Winds. I.Terminal Wind Velocity}
\newblock {\it Acta Astronomica} {\bf 52}, 81-104 (2002) 

\bibitem[17]{leonard00} 
{{Leonard}, D.~C., {Filippenko}, A.~V., {Barth}, A.~J., {Matheson}, T.}
{Evidence for Asphericity in the Type IIn Supernova SN 1998S.}
\newblock {\it \apj} {\bf 536}, 239-254 (2000) 

\bibitem[18]{hillier87} 
{{Hillier}, D.~J.}
{An empirical model for the Wolf-Rayet star HD 50896.}
\newblock {\it \apjs} {\bf 63}, 965-981 (1987) 

\bibitem[19]{ofek13} 
{{Ofek}, E.~O., {Lin}, L., {Kouveliotou}, C., {Younes}, G., {G{\"o}{\v g}{\"u}{\c s}}, E., {Kasliwal}, M.~M., et al.}
{SN 2009ip: Constraints on the Progenitor Mass-loss Rate}
\newblock {\it \apj} {\bf 768}, 47 (2013) 

\bibitem[20]{chugai09} 
{{Chugai}, N.~N.}
{Circumstellar interaction in type Ibn supernovae and SN 2006jc}
\newblock {\it \mnras} {\bf 400}, 866-874 (2009) 

\bibitem[21]{hachinger12} 
{{Hachinger}, S., {Mazzali}, P.~A., {Taubenberger}, S., {Hillebrandt}, W., {Nomoto}, K., {Sauer}, D.~N.}
{How much H and He is `hidden' in SNe Ib/c? - I. Low-mass objects.}
\newblock {\it \mnras} {\bf 422}, 70-88 (2013) 

\bibitem[22]{vink11} 
{{Vink}, J.~S., {Muijres}, L.~E., {Anthonisse}, B., {de Koter}, A., {Gr{\"a}fener}, G. and {Langer}, N.}
{Wind modelling of very massive stars up to 300 solar masses}
\newblock {\it \aap} {\bf 531}, A132 (2013) 

\bibitem[23]{graefener12} 
{{Gr{\"a}fener}, G., {Owocki}, S.~P. and {Vink}, J.~S.}
{Stellar envelope inflation near the Eddington limit. Implications for the radii of Wolf-Rayet stars and luminous blue variables}
\newblock {\it \aap} {\bf 538}, A40 (2012) 

\bibitem[24]{crockett08} 
{{Crockett}, R.~M., {Eldridge}, J.~J., {Smartt}, S.~J., {Pastorello}, A., {Gal-Yam}, A., {Fox}, D.~B., et al.}
{The type IIb SN 2008ax: the nature of the progenitor}
\newblock {\it \mnras} {\bf 391}, L5-L9 (2008) 

\bibitem[25]{arcavi11} 
{{Arcavi}, I., {Gal-Yam}, A., {Yaron}, O., {Sternberg}, A., {Rabinak}, I., {Waxman}, E., et al.}
{SN 2011dh: Discovery of a Type IIb Supernova from a Compact Progenitor in the Nearby Galaxy M51}
\newblock {\it \apjl} {\bf 742}, L18 (2011) 

\bibitem[26]{bersten12} 
{{Bersten}, M.~C., {Benvenuto}, O.~G., {Nomoto}, K., {Ergon}, M., {Folatelli}, G., {Sollerman}, J., et al.}
{The Type IIb Supernova 2011dh from a Supergiant Progenitor}
\newblock {\it \apj} {\bf 757}, 31 (2011) 

\bibitem[27]{vandyk13} 
{{Van Dyk}, S.~D., {Zheng}, W., {Clubb}, K.~I., {Filippenko}, A.~V., {Cenko}, S.~B., {Smith}, N., et al.}
{The Progenitor of Supernova 2011dh has Vanished}
\newblock {\it \apjl} {\bf 772}, L32 (2011) 

\bibitem[28]{ergon13} 
{{Ergon}, M., {Sollerman}, J., {Fraser}, M., {Pastorello}, A., {Taubenberger}, S., {Elias-Rosa}, N., et al.}
{Optical and near-infrared observations of SN 2011dh - The first 100 days}
\newblock {\it \aap} {\bf in press},  (2013) 

\bibitem[29]{arcavi10} 
{{Arcavi}, I., {Gal-Yam}, A., {Kasliwal}, M.~M., {Quimby}, R.~M., {Ofek}, E.~O., {Kulkarni}, S.~R., et al.}
{Core-collapse Supernovae from the Palomar Transient Factory: Indications for a Different Population in Dwarf Galaxies.}
\newblock {\it \apj} {\bf 721}, 777-784 (2010) 

\bibitem[30]{yoon12} 
{{Yoon}, S.-C., {Gr{\"a}fener}, G., {Vink}, J.~S., {Kozyreva}, A. and {Izzard}, R.~G.}
{On the nature and detectability of Type Ib/c supernova progenitors}
\newblock {\it \aap} {\bf 544}, L11 (2012) 


\end{thebibliography}
